\documentclass[letterpaper,11pt]{article}
\usepackage[T1]{fontenc}
\usepackage[utf8]{inputenc}

\usepackage{graphicx}
\usepackage{epstopdf}
\usepackage{multicol}
\usepackage{lmodern}

\usepackage{jheppub}
\usepackage{microtype}   % Finetuned typesetting control
\usepackage{multirow, amsthm}
\usepackage{braket}
\usepackage{mathrsfs}
\usepackage{tikz}
\usepackage[caption=false]{subfig}
\usepackage{xspace}
\usepackage{xcolor}
\usepackage{float,color}
\usepackage[separate-uncertainty=true]{siunitx}
\usepackage[ruled,norelsize]{algorithm2e}
\usepackage[countmax]{subfloat}
\usepackage{accents}
\usepackage{physics}
\usepackage[capitalise]{cleveref}    % Smart referencing

\newcommand{\pythia}[1]{\textsc{Pythia\xspace #1}}

\usepackage{tikz}
\usetikzlibrary{arrows,shapes}
\usetikzlibrary{trees}
\usetikzlibrary{matrix,arrows} 				% For commutative diagram
\usetikzlibrary{positioning}				% For "above of=" commands
\usetikzlibrary{calc,through}				% For coordinates
\usetikzlibrary{decorations.pathreplacing}  % For curly braces
\usepackage{pgffor}							% For repeating patterns

\usetikzlibrary{decorations.pathmorphing}	% For Feynman Diagrams
\usetikzlibrary{decorations.markings}
\usetikzlibrary{snakes}
\tikzset{
    vector/.style={decorate, decoration={snake}, draw},
	provector/.style={decorate, decoration={snake,amplitude=2.5pt}, draw},
	antivector/.style={decorate, decoration={snake,amplitude=-2.5pt}, draw},
    fermion/.style={draw=black, postaction={decorate},
        decoration={markings,mark=at position .55 with {\arrow[draw=black]{>}}}},
    fermionbar/.style={draw=black, postaction={decorate},
        decoration={markings,mark=at position .55 with {\arrow[draw=black]{<}}}},
    fermionnoarrow/.style={draw=black},
    gluon/.style={decorate, draw=black,
        decoration={coil,amplitude=4pt, segment length=5pt}},
    scalar/.style={dashed,draw=black, postaction={decorate},
        decoration={markings,mark=at position .55 with {\arrow[draw=black]{>}}}},
    scalarbar/.style={dashed,draw=black, postaction={decorate},
        decoration={markings,mark=at position .55 with {\arrow[draw=black]{<}}}},
    scalarnoarrow/.style={dashed,draw=black},
    electron/.style={draw=black, postaction={decorate},
        decoration={markings,mark=at position .55 with {\arrow[draw=black]{>}}}},
	bigvector/.style={decorate, decoration={snake,amplitude=4pt}, draw},
}

\tikzstyle{block} = [draw, rectangle, 
    minimum height=3em, minimum width=6em]

\newcommand{\delphes}[1]{\textsc{Delphes\xspace #1}}

\newlength{\dhatheight}

\providecommand{\href}[2]{#2}
\usepackage{comment}

\definecolor{darkred}{rgb}{0.5,0.0,0.0}
\definecolor{darkblue}{rgb}{0.0,0.0,0.9}
\definecolor{darkerblue}{rgb}{0.0,0.0,0.5}
\definecolor{darkgreen}{rgb}{0.0,0.5,0.0}
\definecolor{black}{rgb}{0.0,0.0,0.0}
\definecolor{brown}{rgb}{0.6,0.4,0.2}

\DeclareSIUnit{\nb}{\nano\barn}
\DeclareSIUnit{\pb}{\pico\barn}
\DeclareSIUnit{\fb}{\femto\barn}
\DeclareSIUnit{\year}{yr}

\title{\boldmath Machine Learning Templates for QCD Factorization in the Search for Physics Beyond the Standard Model
}
\author[1]{Joshua Lin,}
\author[2]{Wahid Bhimji,}
\author[3]{and Benjamin Nachman}
\affiliation[1]{\normalsize Department of Physics, University of California, Berkeley, Berkeley, CA 94720, USA}
\affiliation[2]{National Energy Research Scientific Computing Center, Berkeley, CA 94720, USA}
\affiliation[3]{\normalsize Physics Division, Lawrence Berkeley National Laboratory, Berkeley, CA 94720, USA}

\emailAdd{joshua.z.lin@berkeley.edu}
\emailAdd{wbhimji@lbl.gov}
\emailAdd{bpnachman@lbl.gov}

\abstract{
High-multiplicity all-hadronic final states are an important, but difficult final state for searching for physics beyond the Standard Model.  A powerful search method is to look for large jets with accidental substructure due to multiple hard partons falling within a single jet.  One way for estimating the background in this search is to exploit an approximate factorization in quantum chromodynamics whereby the jet mass distribution is determined only by its kinematic properties.  Traditionally, this approach has been executed using histograms constructed in a background-rich region.  We propose a new approach based on Generative Adversarial Networks (GANs).  These neural network approaches are naturally unbinned and can be readily conditioned on multiple jet properties.  In addition to using vanilla GANs for this purpose, a modification to the traditional WGAN approach has been investigated where weight clipping is replaced by drawing weights from a naturally compact set (in this case, the circle).  Both the vanilla and modified WGAN approaches significantly outperform the histogram method, especially when modeling the dependence on features not used in the histogram construction.   These results can be useful for enhancing the sensitivity of LHC searches to high-multiplicity final states involving many quarks and gluons and serve as a useful benchmark where GANs may have immediate benefit to the HEP community.

}

\begin{document} 
\maketitle
\flushbottom

%%%%%%%%%%%%%%%%%%%%%%
\section{Introduction}
\label{sec:intro}
%%%%%%%%%%%%%%%%%%%%%%

Even though collimated sprays of particles (jets) produced from high energy quarks and gluons are ubiquitous at the Large Hadron Collider (LHC), analyzing their substructure has proven to be a powerful tool in the search for physics beyond the Standard Model (SM)~\cite{Larkoski:2017jix,Asquith:2018igt}.  Many theories of physics beyond the SM predict new particles with cascade decays that can result in large multiplicity final states.  When many quarks and gluons are produced in these cascades, multiple large-radius jets with non-trivial substructure can be created~\cite{Cohen:2012yc,Hedri:2013pvl,Hook:2012fd}.  As a result, one powerful method for searching for new particles in the all-hadronic channel is to look for events with a large $\sum_{j\in J} m_j$, where $m_j$ is the jet mass and $J$ is a set of jets in an event.  The key challenge for such an analysis is to estimate the SM background, as high multiplicity multi-jet final states are difficult to accurately predict with current simulation tools.  

Based on the approximate factorization of quantum chromodynamic (QCD) jet production at the LHC~\cite{Collins:1989gx}, the authors of Ref.~\cite{Cohen:2014epa} proposed an innovative background estimation technique.  The idea of the procedure is to estimate the conditional probability $p(m_j|\text{jet kinematics})$ with an event selection suppressed in signal and then to convolve it with the jet kinematic spectrum in the signal region (from data).  A comparison between the predicted $m_j$ and observed $m_j$ is then sensitive to the presence of new particles.  The ATLAS collaboration has successfully applied this method in both Run 1 and Run 2 to set strong limits on potential gluino and squark production~\cite{Aad:2015lea,Aaboud:2018lpl}.

The background estimation procedure described above has two major limitations\footnote{The extensive smoothing studies in Ref.~\cite{Cohen:2014epa} help to mitigate binning effects, but do not have an impact on the feature conditioning challenge.}.  First, $p(m_j|\text{jet kinematics})$ is represented as a histogram and each bin is uncorrelated so many events are needed for a precise determination.  Second, there are physics and detector effects which change the distribution of the jet mass between the region it is constructed (`trained') and the region where it is applied (`tested').  For example, the quark/gluon composition of the background may be different between the two regions.  One way to mitigate this source of method bias is to condition on more features of the jet when constructing the conditional probability.  To reduce the impact of changes in quark/gluon composition, one could add the number of charged-particle tracks inside the jet.  Gluon jets tend to have more particles than quark jets due to their larger color factor.  Since the templates $p(m_j|\text{jet kinematics})$ are binned, it is not simple to condition on more features as one needs many more bins and thus larger samples of events for training.

%Recent studies in High Energy Physics have seen Machine Learning algorithms becoming a popular candidate for the statistical analysis of data. For the problem of classification, 

This paper proposes a solution to both of these limitations using modern machine learning.  Deep neural networks are becoming popular tools for classification and regression tasks in high energy physics (HEP) data analysis, but there is a growing machine learning literature on neural network-based generative models as well.  Training a generative model can be viewed as a regression task that maps noise to structure, mimicking the Jacobian from a pre-defined probability distribution to a target probability distribution.  One of the most well-studied paradigms for such models is the Generative Adversarial Network (GAN)\cite{Goodfellow:2014upx} (details in Sec.~\ref{sec:architectures}).  GANs have also been studied in HEP and show great promise for accelerating simulations~\cite{Paganini:2017hrr,Paganini:2017dwg,deOliveira:2017rwa,Chekalina:2018hxi,Carminati:2018khv,Vallecorsa:2018zco,Erdmann:2018jxd,Musella:2018rdi,Erdmann:2018kuh,ATL-SOFT-PUB-2018-001,deOliveira:2017pjk} and may also be useful for other tasks such as sampling from the space of effective field theory models~\cite{Erbin:2018csv}.  In the context of QCD factorization studied in this paper, the GAN will learn the probability distribution of the jet mass given the jet kinematics and any other useful information.

This paper is organized as follows.   Section~\ref{sec:architectures} introduces GANs and how they can be used to exploit QCD factorization.   The application of GANs to the phase space relevant to the Supersymmetry (SUSY) search from Ref.~\cite{Cohen:2014epa,Aad:2015lea,Aaboud:2018lpl} is presented in Sec.~\ref{sec:templates}.  The paper ends with conclusions and outlook in Sec.~\ref{sec:conclusions}.

%%%%%%%%%%%%%%%%%%%%%%
\section{Machine Learning Architectures}
\label{sec:architectures}
%%%%%%%%%%%%%%%%%%%%%%

%%%%%%%%%%%%%%%%%%%%%%
\subsection{Overview of GAN setup}
\label{sec:vanillagan}
%%%%%%%%%%%%%%%%%%%%%%

The goal of this section is to introduce an approach to learn the conditional distribution of the jet mass given various kinematic features.  Neural networks are chosen due to their flexibility and the resulting algorithms are naturally unbinned.  There are multiple neural network-based approaches to generative modeling such as Variational Autoencoders (VAE)~\cite{DBLP:journals/corr/KingmaW13,Rezende:2014:SBA:3044805.3045035}, Mixture Density Networks (MDN)~\cite{mdn}, and Generative Adversarial Networks (GAN)~\cite{Goodfellow:2014upx}.  GANs are selected because they can readily model asymmetric distributions and accommodate conditional features.  

%Our neural networks were designed to be capable of learning the mass distributions of jets from their kinematic variables in a bin-less manner. To learn these templates, we designed Generative Adversarial Network (GAN) architectures, as first described in \cite{Goodfellow:2014upx}. In contrast to usual applications of GANs, such as generation of images by learning the entire distribution of the training data set; our GANs aim to learn the \textit{conditional} mass distributions (templates $p(m_j | \text{ jet kinematics})$) of the mass given the kinetic variables; in order to dress the jet masses in our final analysis. 
%Our GANs slightly differ from regular GANs in the sense that regular GANs aim to learn the entire distribution, 

\begin{figure}
\begin{center}
\includegraphics[width=\textwidth]{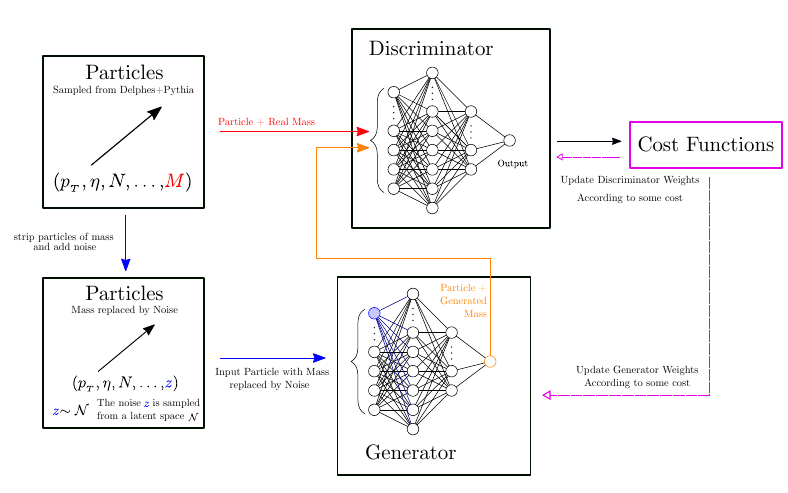}
\end{center}
\caption{Flowchart describing how GANs are used to learn templates (shown here mass templates for the RPV-SUSY search) given kinematic variables. The generator network is a feed-forward network that takes as input particles with their mass replaced by noise, and generates mass according to a learnt distribution. These fake particles are bundled with real particles and passed to the discriminator, which learns to discriminate between the real and fake distributions. }
\label{fig:GAN_overview}
\end{figure}

Generative Adversarial Network training uses a pair of neural networks: one to map noise into structure (generator) and one to classify (discriminator) physics-based examples from the generator examples.  The generator is structured as a densely connected feed-forward neural network that takes as input both jet kinematic features and noise and outputs a jet mass.  The generated masses are then input to the discriminator network where they are compared against masses from a physics-based generator that match the kinematic quantities.  The two networks `compete' until the discriminator network is as bad as possible, which means that the generator is proficient at modeling the conditional jet mass distribution.  This minimax structure uses the loss function described in Ref.~\cite{Goodfellow:2014upx}, constructed to minimize the Jensen-Shannon divergence between the distribution of the real data (in this case, a physics-based simulator) and the distribution of the generated data.  A schematic of this GAN is shown in \cref{fig:GAN_overview}. 

%To learn the conditional distributions, our GAN (which consists of dense feed-forward layers) takes as input both kinematic variables and noise, and outputs generated masses for the kinematic variables provided. The kinematic variables and generated masses are then compared against the kinematic variables with the real masses by the discriminator; and the two neural networks are appropriately updated by the loss function described in \cite{Goodfellow:2014upx}, constructed to minimize the Jensen-Shannon divergence between the distribution of the real data and the distribution of the generated data.  A schematic of this GAN is shown in \cref{fig:GAN_overview}.  

The GAN networks for the jet mass are relatively small compared to others in the literature, which are mostly used for modeling image data.  For the vanilla GAN implementation, both the Generator and Discriminator networks are composed of three hidden layers between input (kinetic variables and noise/mass for generator/discriminator respectively) and output (generated mass/likelihood for generator/discriminator respectively).  The first layer has 160 neurons, the second has 80 neurons, and the last hidden layer has 40 neurons.  Additionally, the generator network is made more robust by adding 50\% dropout~\cite{JMLR:v15:srivastava14a}.  Network weights were chosen using the Adam optimizer~\cite{DBLP:journals/corr/KingmaB14} with early stopping.  These settings were chosen after a modest hyper-parameter scan.   

%Our basic GAN implementation follows that of the original GAN paper, with loss given by that of the paper. 
%Our networks are relatively small compared to others seen in the literature, as in our context the phase space of jets has low dimension compared to say the space of images. For the ordinary GAN implementation, both the Generator and Discriminator are comprised of three intemediary layers between input (kinetic variables and noise/mass for generator/discriminator respectively) and output (generated mass/likelihood for generator/discriminator respectively), the first with 160 neurons, then 80 neurons, then 40 neurons. We found that using dropout\cite{JMLR:v15:srivastava14a} in the Generator network (of 50\%) improved results. We used the ADAM optimizer\cite{DBLP:journals/corr/KingmaB14}, and for regularization used the Early Stopping Method. 
%Say something about testing different models and optimizing to arrive at this.

All of the neural networks are built using \textsc{Tensorflow}~\cite{tensorflow} on Nvidia GeForce 1080 Ti GPUs.  Since the jet mass distribution is skewed toward the high mass tail, the input noise distribution was varied according to a skew-normal distribution~\cite{skewnormal}: $f(x|\alpha)=2\phi(x)\Phi(\alpha x)$, where $\alpha$ is a hyper-parameter and $\phi$ and $\Phi$ are the probability density and cumulative distribution of the standard normal distribution, respectively.  The best value of the skew parameter was identified to be $\alpha=5$ (labeled \textit{skew} in the results plots).  This will be compared to the standard no-skew case ($\alpha=0$; labeled \textit{noskew}).

 %We also found that by using a skewed distribution for the prior input random distribution to the generator network helped the GAN match the skew of the mass distribution in the high mass limit. Specifically, we tested using a Gaussian distribution for the prior random distribution (labelled noskew in our plots) against using a skew-normal distribution with skew $\alpha = 5$ (labelled skew in our plots).
%{\color{blue}(Read more into the literature about changing up the noise generator?)}
%'Our training was performed using the Keras Python neural network library with Tensorflow backend, on
%Nvidia GeForce 1080 Ti GPUs' %Sentence taken from previous article; make it different?
%https://arxiv.org/pdf/1611.07004v1.pdf

%%%%%%%%%%%%%%%%%%%%%%
\subsection{Modified WGAN}
%%%%%%%%%%%%%%%%%%%%%%

As an alternative to the vanilla GAN described in Sec.~\ref{sec:vanillagan}, a popular variant called the Wasserstein-GAN (WGAN)~\cite{1701.07875} was also studied (for another HEP application, see Ref.~\cite{Erdmann:2018jxd}). The WGAN differs from the vanilla GAN in that it minimizes the Earth-Mover distance (also known as the Wasserstein Distance):

\begin{align}
W(\mathbb{P}_\text{real},\mathbb{P}_\text{generated}) = \inf_{\gamma \in \Pi (\mathbb{P}_\text{real},\mathbb{P}_\text{generated})} \mathbb{E}_{(x,y) \sim \gamma} [ \parallel x - y \parallel ],
\end{align}

\noindent where $\gamma\in \Pi (\mathbb{P}_\text{real},\mathbb{P}_\text{generated})$ is a joint distribution with marginal distributions $\mathbb{P}_\text{real},\mathbb{P}_\text{generated}$ respectively, and $(x,y) \sim \gamma$ means that the random variable $(x,y)$ is drawn from the distribution $\gamma$. This 'softer' metric was introduced as a way to combat the vanishing gradients that often occured when training regular GANs \cite{1701.07875,Arjovsky2017TowardsPM}. In the algorithm suggested by the WGAN paper to minimise the Earth-Mover distance, the discriminator is a function $f$ that is taken from a space of trial functions that are all $K$-Lipshitz for some $K$. To enforce such conditions, the functions $f$ are constructed as feed-forward neural networks with weights $w$ that are clipped (after every update) to a compact space $[-\alpha,\alpha]$ for some fixed $\alpha$ which enforces $K$-lipshitz for some $K$. 

An exact implementation of the WGAN approach resulted in weights that would often aggregate around the specific limit values $\alpha$ chosen, leading to vanishing gradients and generated mass distributions that did not match the physical mass distributions. The challenges surrounding the clipping operation to enforce the Lipshitz condition are discussed in the original WGAN paper, and explored in variations of WGAN where the clipping is replaced by gradient penalty~\cite{DBLP:journals/corr/GulrajaniAADC17} or asymmetric clipping~\cite{2017arXiv170507164G}. 

\begin{figure}
\begin{center}
\includegraphics[scale=2.25]{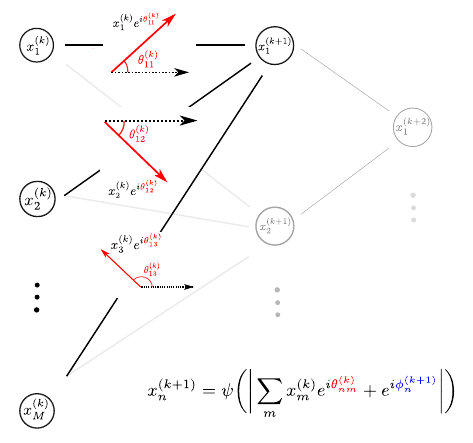}
\end{center}
\caption{A schematic diagram to illstrate the modified WGAN, where all the weights and biases are angles.  The compactness of the unit circle then guarantees that the trained functions are all $K$-lipschitz for sufficiently large $K$. The inputs to the layer are $x_1,...,x_n$ and there is a hidden layer with two nodes $y_1$ and $y_2$ (in general, there can be many more than two) followed by a single output layer $z_1$.  The equation shows the form of the activation that maps $x_1,...,x_n$ to $y_1$.  }
%\textbf{See the text for how I modified the notation on the activation - would be good to harmonize with the figure.}}
\label{fig:SGAN}
\end{figure}

In this paper, a modification to weight clipping is introduced that changes how the weights act on the outputs of neurons to enforce the Lipshitz condition. A schematic for this modification to the WGAN is shown in \cref{fig:SGAN}. The original weight-clipping operation enforces a Lipshitz function because the composition of Lipshitz functions is still a Lipshitz function; in particular, there is the weak assumption that all the activation functions used in the neural networks are themselves Lipshitz (true for the most popular activation functions including sigmoid, tanh, and ReLU). Then, the fact that there is a single constant $K$ for which all the functions $f$ that we are considering are $K$-Lipshitz is due to the weight-clipping -- this restricts the function space to be compact. 

Another natural way to enforce the Lipshitz constraint that eliminates boundary effects is to draw weights from a compact space with no boundaries. One way to achieve this is to draw the weights and biases from the unit circle\footnote{In principle, one could generalize this idea to modify the WGAN where all weights and biases are drawn from a generic compact Lie Group.  The outputs of the neurons exist in a vector space with the Lie Group acting by a chosen representation. In this framework, a normal linear neural network uses the Lie Group $\mathbb{R}$ with canonical action on itself, and the modified WGAN shown in Eq.~\ref{eq:modifiedwgan} uses $U(1) \approx S^1$. The point of this construction is that the compactness of the weight-space assures the convergence of the WGAN and is well-formed because Lie Group actions can be differentiated. Such applications of Lie Groups to Machine Learning have been explored elsewhere in the literature, for example in applications to 3D-classification problems~\cite{DBLP:journals/corr/HuangWPG16}.}:

\begin{align}
\label{eq:modifiedwgan}
x^{(k+1)}_n = \psi \bigg(\bigg|\sum_m x^{(k)}_m e^{i \theta^{(k)}_{nm}} + e^{i \phi^{(k+1)}_n}\bigg|\bigg),
\end{align}

\noindent where $x^{(k)}_m$ represents the output of the $m$-th neuron in the $k$-th layer of the neural network and $\psi$ is the activation function.  The bias is replaced by $b_i=e^{i\phi_n}$ and the weights are replaced by $w_{nm}=e^{i\theta_{nm}^{(k)}}$.  %A schematic for this modification to the WGAN is shown in \cref{fig:SGAN}.

%Note that although weight clipping is one method in which to enforce the $K$-lipshitz property for all the functions $f$ represented by the neural network; in actuality it is sufficient for the space of functions $\mathcal{F}$ that we are drawing our critic function $f$ from to be compact, for the functions to satisfy the $K$-lipshitz property. 

%In this paper, we introduce a modification to the WGAN (which shows better performance than the regular WGAN as described in the seminal paper), that replaces the 'weight clipping' operation introduced to enforce a finite maximum Lipshitz constant for the functions in the space of functions we are training over. The 'weight clipping' operation clips all the weights of the neural network to a limited range around $0$, (for example to the interval $[-0.01,0.01]$) to ensure that the functions are all $K$-lipshitz for some large $K$. When implementing this approach, we faced the issue that many times our weights would aggregate around the limit values (in our example, $\pm 0.01$) and the neural network would suffer from effective vanishing gradient due to the weight-clipping. Instead of clipping the weights, we found that treating the weights and biases of the feedforward network to be living in a compact space such as $S^1$ gave better results. Specifically, we can treat the weights and biases as angles, $\theta_i, \phi_i$; and we modify the manner in which our weights are combined as shown in \cref{fig:SGAN}.

\section{Templates for RPV-SUSY}
\label{sec:templates}
%%%%%%%%%%%%%%%%%%%%%%

%%%%%%%%%%%%%%%%%%%%%%
\subsection{Overview of RPV-SUSY and the Template Method}
\label{subsec:RPV-SUSY}
%%%%%%%%%%%%%%%%%%%%%%

%(DRAFT) - {\color{blue}blue text are comments}

\begin{figure}[h!]

\begin{center}
\begin{tikzpicture}[line width=1.5 pt, scale=1.3]
	
	\draw (-1,1+0.1)--(0,0+0.1);
	\draw (-1,-1+0.1)--(0,0+0.1);
	
	\draw (-1,1)--(0,0);
	\draw (-1,-1)--(0,0);
	
	\draw (-1,1-0.1)--(0,0-0.1);
	\draw (-1,-1-0.1)--(0,0-0.1);		
	
	\draw[gluon,color=red] (0,0)--(1,1);
	\draw[gluon,color=red] (0,0)--(1,-1);
	\draw[color=red] (0,0)--(1,1);
	\draw[color=red] (0,0)--(1,-1);

	\draw[dashed,color=red] (1,1)--(2,1.5);
	\draw[dashed,color=red] (1,-1)--(2,-1.5);

	\draw[] (1,1)--(2,0.5);
	\draw[] (1,-1)--(2,-0.5);

	\draw[color=red] (2,1.5)--(3,2);
	\draw[color=red] (2,-1.5)--(3,-2);
	\draw[vector,color=red] (2,1.5)--(3,2);
	\draw[vector,color=red] (2,-1.5)--(3,-2);	
	\draw[] (2,1.5)--(3,1);
	\draw[] (2,-1.5)--(3,-1);
	
	\draw[] (3,-2) -- (4,-2.5);
	\draw[] (3,-2) -- (4,-2);
	\draw[] (3,-2) -- (4,-1.5);

	\draw[] (3,2) -- (4,2.5);
	\draw[] (3,2) -- (4,2);
	\draw[] (3,2) -- (4,1.5);
	
	\draw[fill=black!30!white] (0,0) circle (.25);	
	
	\node at (1.5, 1.6) {\color{red} $\tilde{q}$};
	\node at (1.5, -1.6) {\color{red} $\tilde{q}$};
	\node at (2.5, 2.1) {\color{red} $\tilde{\chi}^0$};
	\node at (2.5, -2.1) {\color{red} $\tilde{\chi}^0$};	
	\node at (0.4, 0.8) {\color{red} $\tilde{g}$};
	\node at (0.4, -0.8) {\color{red} $\tilde{g}$};	
	\node at (2.2, 0.4) {$q$};
	\node at (2.2, -0.4) {$q$};
	\node at (3.4, 0.9) {$q$};
	\node at (3.4, -0.9) {$q$};

	\node at (4.4, -2.5) {$q$};
	\node at (4.4, -2.) {$q$};
	\node at (4.4, -1.5) {$q$};

	\node at (4.4, 2.5) {$q$};
	\node at (4.4, 2.) {$q$};
	\node at (4.4, 1.5) {$q$};
		
 \end{tikzpicture}
\end{center}
\caption{Schematic Feynman-like diagram for RPV SUSY.}
\label{fig:feynman}
\end{figure}
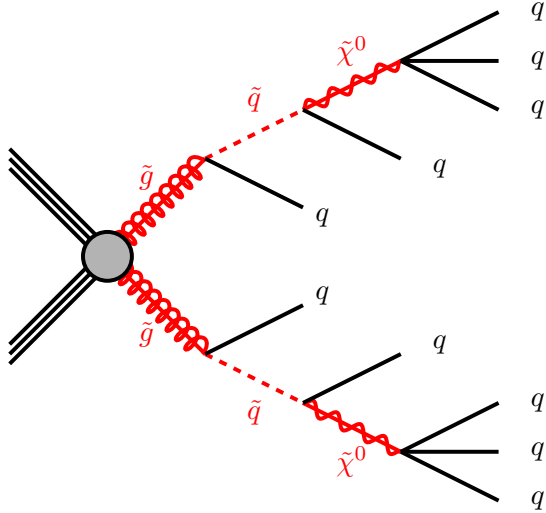

Supersymmetry (SUSY)~\cite{Golfand:1971iw,Volkov:1973ix,Wess:1974tw,Wess:1974jb,Ferrara:1974pu,Salam:1974ig} is a well-studied extension of the Standard Model in which there is a new symmetry relating fermions and bosons.  In models of SUSY that conserve $R$-parity~\cite{FARRAR1978575,DIMOPOULOS1981150,PhysRevD.26.287,SAKAI1982533,DIMOPOULOS1982133} -- an additional symmetry that requires SUSY particles to be produced in pairs -- collider signatures often feature large missing transverse momentum (MET) carried by the lightest supersymmetric particle (LSP) which must be neutral under the electromagnetic and strong forces.  However, $R$-parity is not present in all SUSY models and collider-based limits relying on MET are typically insensitive to such models.  One $R$-parity violating (RPV) coupling (often denoted $\lambda''$) results in gluino/neutralino decay into three quarks, as illustrated in Fig.~\ref{fig:feynman}.  When this coupling is present, traditional searches for SUSY are largely ineffective because there can be little MET and no charged leptons.  Despite lacking standard handles for separating potential SUSY events from SM background processes, RPV signatures like those in Fig.~\ref{fig:feynman}, all-hadronic SUSY searches have been able to set strong limits on gluino and squark production in models with a large $\lambda''$ (see references within Ref~\cite{Aad:2015lea,Aaboud:2018lpl} for other constraints on such models). 

%, supersymmetric partners are introduced for the Standard Model particles, introducing a wide array of additional interactions. A symmetry of interest between SM particles and their superpartners is $R$-parity; but in generality is not necessarily conserved in SUSY models. In particular; the $R$-partity violating (RPV) terms of the superpotential for the MSSM can be generically written as \cite{Dreiner:1997uz}:
%$$W_\text{RPV} = \frac{1}{2} \lambda_{ijk} L_i L_j \bar{E}_k + \lambda'_{ijk} L_i Q_j \bar{D}_k + \frac{1}{2} \lambda''_{ijk}\bar{U}_i \bar{D}_j \bar{D}_k + \kappa_i L_i H_2$$
%These dimension 3 superpotential terms lead to dimension 4 terms in the RPV Lagrangian. Specifically, the $\lambda''_{ijk}$ term (also called the $UDD$ term) corresponds to gluino/neutralino decay into three quarks, such as the case illustrated in Fig.~\ref{fig:feynman}.   

One approach to search for all-hadronic SUSY decays like those in Fig.~\ref{fig:feynman} is to exploit the high-multiplicity of hard, well-separated partons in the final state.  In the recent ATLAS analysis~\cite{Aaboud:2018lpl}, the main kinematic observable used in the search for $\lambda''$ RPV decays is the total jet mass $M_\Sigma^J$ defined as the sum of the masses of the four leading large $R=1.0$ jets.  Multiple jets with a large mass are generated from well-separated high-energy partons that happen to be clustered within a single jet.  The SM multijet background is estimated by a data-driven method~\cite{Cohen:2014epa}, whereby mass templates are constructed from control regions containing a low purity of potential signal.  These templates are histograms that model the dependence of QCD mass on jet kinematic properties ($p_T$ and $\eta$).  If the jets in the signal region are only due to QCD, then the mass distribution can be estimated by convolving the jet kinematics with the mass templates.  This estimate is compared with the actual mass distribution and deviations would be an indication of BSM physics.

The goal of this study is to demonstrate that GANs may be a useful alternative method to simple histogram templates for learning the dependence of the jet mass on jet properties.  

%In this study, our goal is to demonstrate that GANs, when correctly applied, are a viable alternative to mass templates to learn the mass distribution of fat jets in the control region given their kinematic properties such as $p_T$ and $|\eta|$; and have the advantage that they can be easily generalizable to train on more variables of interest such as neutral track multiplicity. To this end, we will be training our GANs in the control region and testing them on the Validation region events. 

%; splitting the phasespace into control, validation, and signal regions. Mass templates, which are histograms modelling the mass distribution as a function of $pT$ and $\eta$ are built from the control region data, validated in the validation region; and tested to see for significant difference in the signal region. 

%%%%%%%%%%%%%%%%%%%%%%
\subsection{Simulation Setup}
\label{subsec:setup}
%%%%%%%%%%%%%%%%%%%%%%

To demonstrate the potential of GAN-based templates, $pp\rightarrow\text{jets}$ at $\sqrt{s} = 13 \text{ TeV}$ are generated with \pythia{8.223}~\cite{Sjostrand:2006za} using the fast detector simulation \delphes{3.4.0}~\cite{deFavereau:2013fsa} with the detector card {\tt delphes\_card\_ATLAS.tcl}. Following the ATLAS analysis strategy~\cite{Aaboud:2018lpl}, events are clustered~\cite{Cacciari:2011ma} into jets using the anti-$k_t$ algorithm~\cite{Cacciari:2008gp} with radius parameter $R = 1$.  These jets are groomed according to the trimming procedure~\cite{Krohn:2009th} where subjets with radius $R=0.2$ are created from the large-radius jet constituents and removed if their transverse momentum is below 5\% of the parent jet's $p_T$.  The remaining large-radius jets are only considered if $p_T>200$ GeV and $|\eta|<2.0$ and are divided into control, validation and signal regions according to \cref{tab:regions}.   The control region is used to construct the templates (train the GAN) and the signal region is where they are applied\footnote{We note that the original template proposal from Ref.~\cite{Cohen:2014epa} called for a smoothing procedure that was used in the earlier ATLAS result~\cite{Aad:2015lea} but not the later one~\cite{Aaboud:2018lpl}.  Our analysis here more closely resembles the later approach.  The smoothing should in principle help mitigate binning effects, though would not change the qualitative conclusions about feature dependence.}.  The validation region is expected to be sufficiently devoid of potential signals that it can be used to study the fidelity of the templates.  Even though the \pythia is only leading order in the strong coupling constant, Ref.~\cite{Aaboud:2018lpl} used the same setup and found a good agreement with data so this setup is sensible for testing new methods.

As a baseline, mass templates (histograms) are constructed following the ATLAS study: jets in the control region divided into $4$ $|\eta|$ bins defined uniformly between $0$ and $2$, and $15$ $p_T$ bins defined uniformly in $\log_{10}(p_T)$. This results in $60$ mass histograms in total.  No $b$-tagging is applied, though future extensions to include flavor tagging information are possible.  Using the mass histograms, jets in the validation and signal regions can be dressed with random masses given their $p_T$ and $\eta$.

%In this study, we focus on the $b$-tag rejected events (by far the majority of the events). For each of the jets, we keep the mass $m$, the $p_T$, $\eta$, and $N$ - the neutral track multiplicity - information. Using the mass templates, we can dress jets in the validation and signal regions with random jet masses. 

\begin{table}
\begin{center}
\begin{tabular}{| c | c | c | c | c |}
\hline
Region & $N_\text{jet} (p_T > 200 \ \text{GeV})$ & $p_{T,1}$ & $|\Delta \eta_{1,2}|$ & $M_J^\Sigma$ \\
\hline 
Control & $=3$ & - & - & - \\
\hline
Validation & $=4$ & $> 400 \ \text{GeV}$ & $> 1.4$ & - \\
 & $\geq 5$ & - & $>1.4$  & - \\
 \hline
 Signal & $=4$ & $> 400 \ \text{GeV}$ & $< 1.4$ & $> 1.0 \ \text{TeV}$ \\
  & $\geq 5$ & - & $<1.4$ & $> 0.8 \ \text{TeV}$ \\
 \hline
\end{tabular}
\caption{Phasespace requirements for the different regions considered.. } 
\label{tab:regions}
\end{center}
\end{table}

%\begin{itemize}
%\item Control Region : 3 Jets
%\item Validation Region : 4 jets , $pT_1 > 400 $ GeV, $|\Delta \eta_{1,2}| > 1.4$, or 5 jets and $|\Delta \eta_{1,2} | > 1.4$. 
%\item Signal Region : 4 jets, $pT_1 > 400 $ GeV, $|\Delta \eta_{1,2}| < 1.4$, $M_J^\Sigma > 1.0 TeV$. 5 jets, $|\Delta \eta_{1,2}| < 1.4$, $M_J^\Sigma > 0.8 TeV$. 
%\end{itemize}

%%%%%%%%%%%%%%%%%%%%%%%
\subsection{Machine Learning Results}
\label{subsec:ML results}
%%%%%%%%%%%%%%%%%%%%%%%

After the selections described in the previous section, there were 1.1 million jets in the control region and 30k in the validation region.  The validation region is used to test the efficacy of the neural network training, filling the role of the `test set' and by construction is independent from the training set.  The jets in the control region are split 50\%-50\% for the purpose of training the neural network and `validating' the network to enforce the early stopping condition.

%After cuts, we had 1.1 million control region fat jets, and 30 thousand validation region fat jets. Note that the validation region jets are actually what we use to test our generated templates on; so for the purposes of neural network training might be called our 'test' dataset. The control region jets were further split 50\% - 50\% for the purposes of training and validating the neural networks. 

The accuracy of the generated mass templates was quantified\footnote{There is no unique way to monitor the GAN performance during training.  For a typical GAN trained with images, this is qualitatively different than classifier training because the entire multi-dimensional probability is being modeled, not just the likelihood ratio.  The 1D case here is not as extreme, but still different than classification or regression monitoring.  One can use the full GAN loss, the discriminator loss, or any divergence that gives a `distance' between probability distributions.  This particular divergence is popular in HEP and is therefore used as a diagnostic here.} using the separation power metric~\cite{Harrison:1998yr,Hocker:2007ht}: 

\begin{align}
\label{eq:separation}
S(\mathbb{P}_1,\mathbb{P}_2) = \frac{1}{2} \int_X \frac{(\mathbb{P}_1(x) - \mathbb{P}_2(x))^2}{\mathbb{P}_1(x) + \mathbb{P}_2(x)} \text{d}x,
\end{align}

%https://arxiv.org/pdf/1810.05653.pdf
\noindent where $\mathbb{P}_1,\mathbb{P}_2$ are probability distributions over a space $X$ and Eq.~\ref{eq:separation} is normalized to be between $0$ and $1$.  For the RPV SUSY case, $X \simeq \mathbb{R}_{p_T} \times \mathbb{R}_\eta \times \mathbb{R}_N \times \mathbb{R}_m$, one real line for each of the jet properties $p_T, \eta$, constituent track multiplicity\footnote{Due to their robustness to pileup and excellent angular resolution, charged-particle tracks are associated to jets and used as proxy for the number of particles inside the jet.} ($N$), and $m$.  The $\mathbb{P}_i$ are the real and generated distributions:

\begin{align}
\label{eq:preal}
\mathbb{P}_\text{real}&(p_T, \eta, N, m)\\
\label{eq:pgen}
 \mathbb{P}_\text{generated}&(p_T,\eta, N,m) = \mathbb{P}_\text{real}(p_T,\eta,N) G (m|p_T,\eta, N),
\end{align}

\noindent where $G (m|p_T,\eta, N)$ is the learned mass distribution by the GAN for a given $p_T,\eta,N$ value, and $\mathbb{P}_\text{real}(p_T,\eta,N) =\int \mathbb{P}_\text{real}(p_T, \eta, N, m)\text{d}m$.  Equation~\ref{eq:pgen} explicitly encodes the QCD factorization of the jet kinematics and the mass given those kinematic properties.  Neither the GAN or physics-based simulator provide $\mathbb{P}$ directly; instead, only examples are drawn from the distributions.  Empirical distributions are constructed from samples and the separation power is approximated by first binning the jets into eight regions of equal statistics in the kinematic variables $p_T, \eta, N$, by splitting the events into two collections based on the jet $p_T$, then splitting each of these collections evenly in $\eta$, and then in $N$.  In each of these eight bins, the jet mass distribution is used to calculate the (binned) separation power for each dataset and then all eight sets\footnote{The estimated separation power, for our dataset size, is not sensitive to increasing bin number beyond $8$.} are combined.

%So as an approximation to the real separation power between the two distributions; we can calculate the separation power of one-dimensional mass-distributions of events in prechosen kinetic phase-space regions, $X_i \subset \mathbb{R}_{p_T} \times \mathbb{R}_\eta \times \mathbb{R}_N$ with $\cup_i X_i = \mathbb{R}_{p_T} \times \mathbb{R}_\eta \times \mathbb{R}_N$ and $X_i \cap X_j$ a null set for $i \neq j$. {\color{blue}(Justification is along the lines of $X_i$ small means error between approximation and true separation power small)} The specific regions $X_i$ chosen correspond to the simple choices of $p_T,\eta,N$ being constrained by equal cuts. In \cref{fig:err}, we show the (approximated) separation power as a function of epoch for various GANs. 

%by choosing dividers $p_{T,1}, p_{T,2}$, $\eta_1,\eta_2$ and $N_1,N_2$ and considering all events with $p_T < p_{T,1}$, $p_{T,1} < p_T < p_{T,2}$, or $p_T > p_{T_2}$, and calculating the separation power of each of the regions, and then summing up these contributions. 

\begin{figure}
\begin{center}
\includegraphics[scale=0.6]{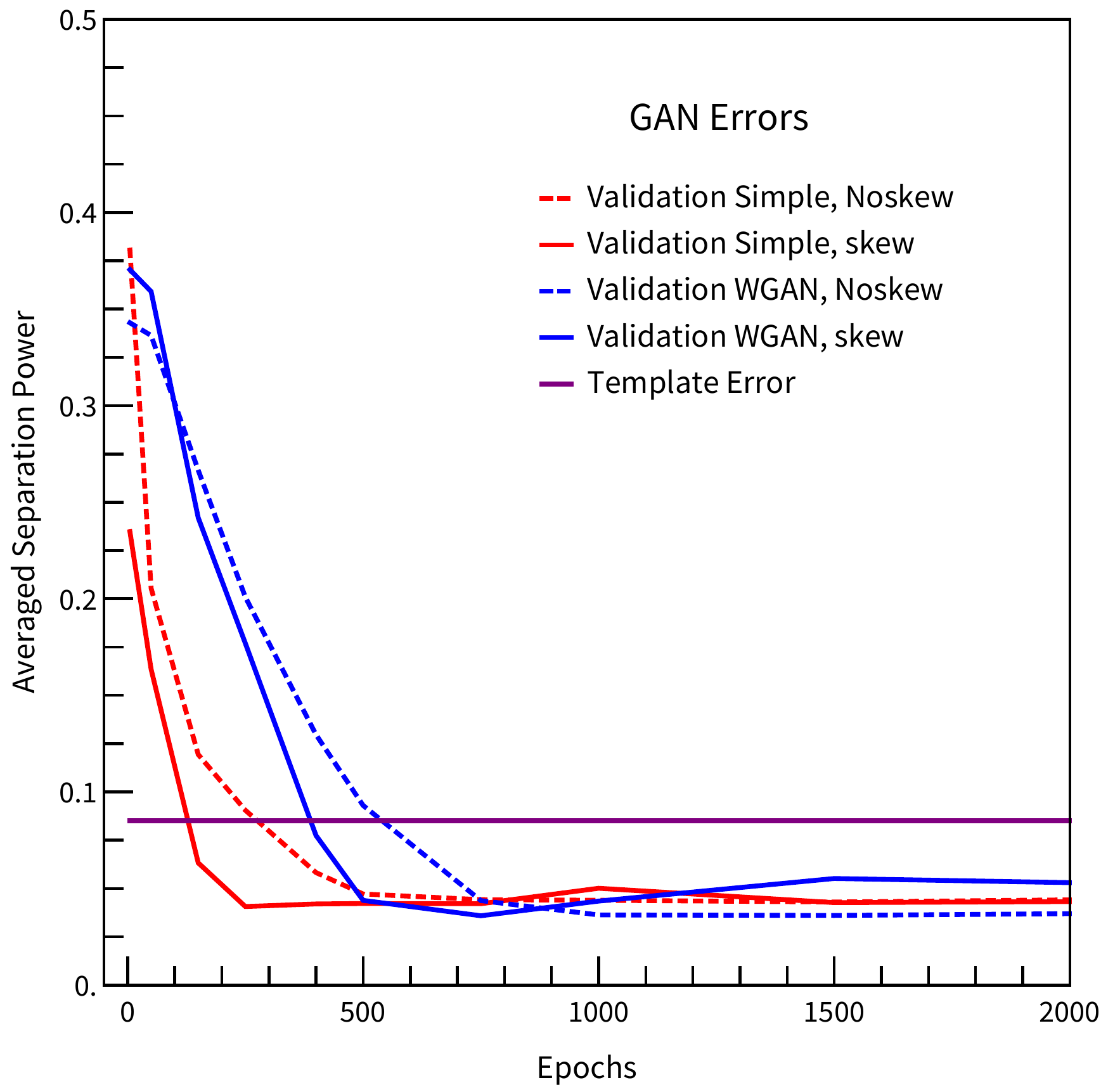}
\end{center}
\caption{Estimated separation power between the generated jet-kinematic distribution and the real jet-kinematic distribution for various GAN architectures.  The template error corresponds to the baseline approach with no neural networks and is thus independent of the number of training epochs.}
\label{fig:err}
\end{figure}

Figure~\ref{fig:err} shows the separation power for various generative models as a function of the number of epochs used to train.  GAN models that were initialized with a high separation error (above $0.6$) training much slower due to vanishing gradients, so only those GANs with an initialized value below 0.6 are considered for Fig.~\ref{fig:err}.  Furthermore, to reduce the impact of fluctuations in the initialization, the average value over ten random initializations are used. By construction, the default template method does not involve neural networks and thus is constant.  As desired, all of the GAN approaches converge to a separation power that is smaller than the template method, as they have access to more and unbinned information.  For both the vanilla GAN and the WGAN, using a skew-normal distribution for the noise accelerates the training time.  The Vanilla GAN also converges faster than the WGAN, though all GAN approaches have a similar final separation power.

\begin{figure}
\vspace{-1.8cm}
\begin{center}
%\includegraphics[width=\textwidth]{figures/hist_ratio.pdf}
%\hbox{\hspace{0cm}\includegraphics[scale=0.55]{figures/hist_ratio.pdf}}
\hbox{\hspace{0cm}\includegraphics[scale=0.55]{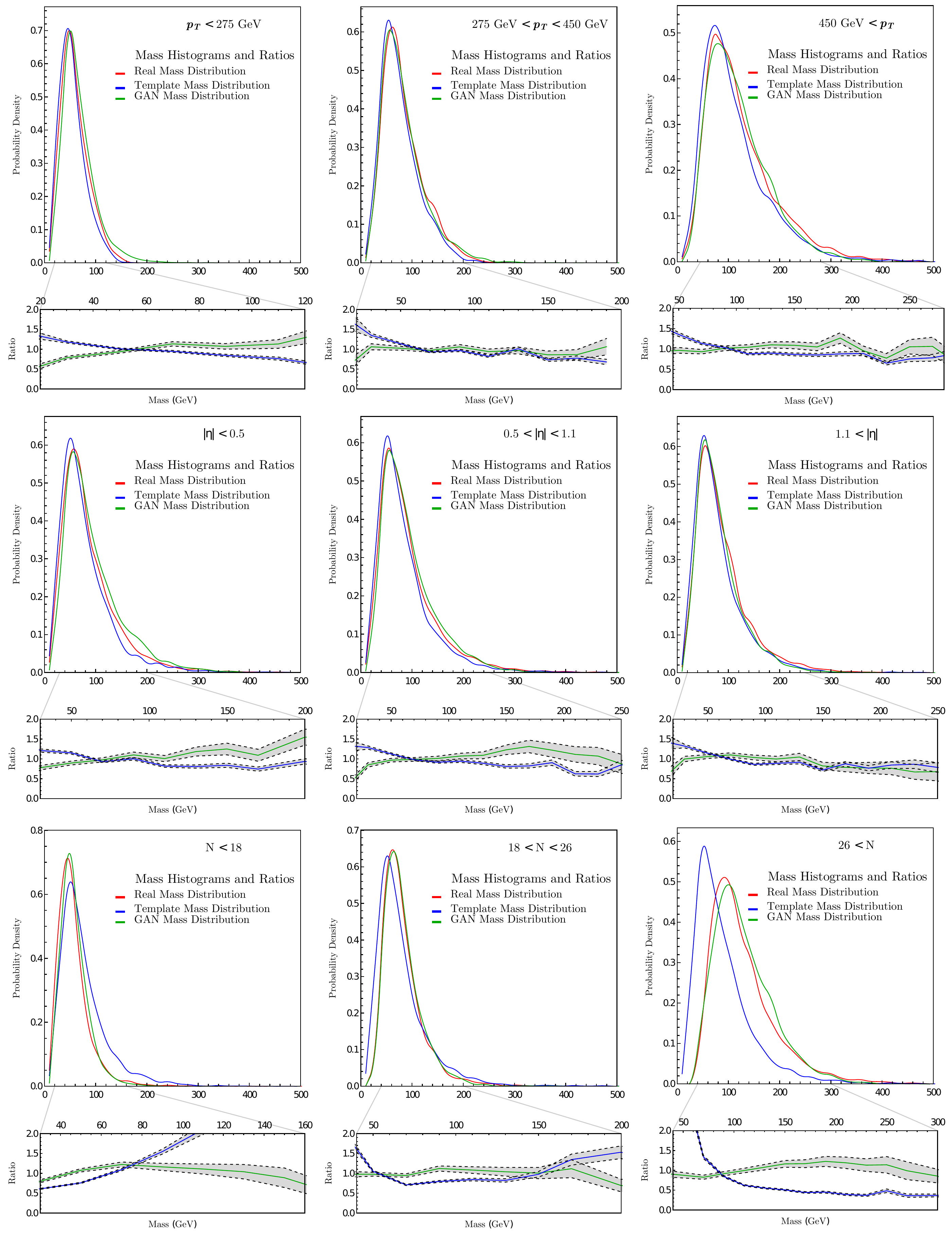}}
\end{center}
\vspace{-0.5cm}
\caption{The physics-based (`real') mass distributions compared with distributions from the template method and the vanilla GAN in bins of jet $p_T$ (top row), $\eta$ (middle row), and $N$ (bottom row). The uncertainty in the ratio was calculated as the $1$-sigma error assuming poisson distributions of events in each bin. The error shown in the plots is the calculated statistical error.  The corresponding plot in the control region is qualitatively similar, but converges quicker.} 
%\textbf{Which GAN is this?  Also, the y-axis should maybe say probability density or something like that?  What is the uncertainty band?  Shouldn't the red ratio line be green?}
\label{fig:ratio}
\end{figure}

\begin{figure}
\begin{center}
\includegraphics[scale=0.31]{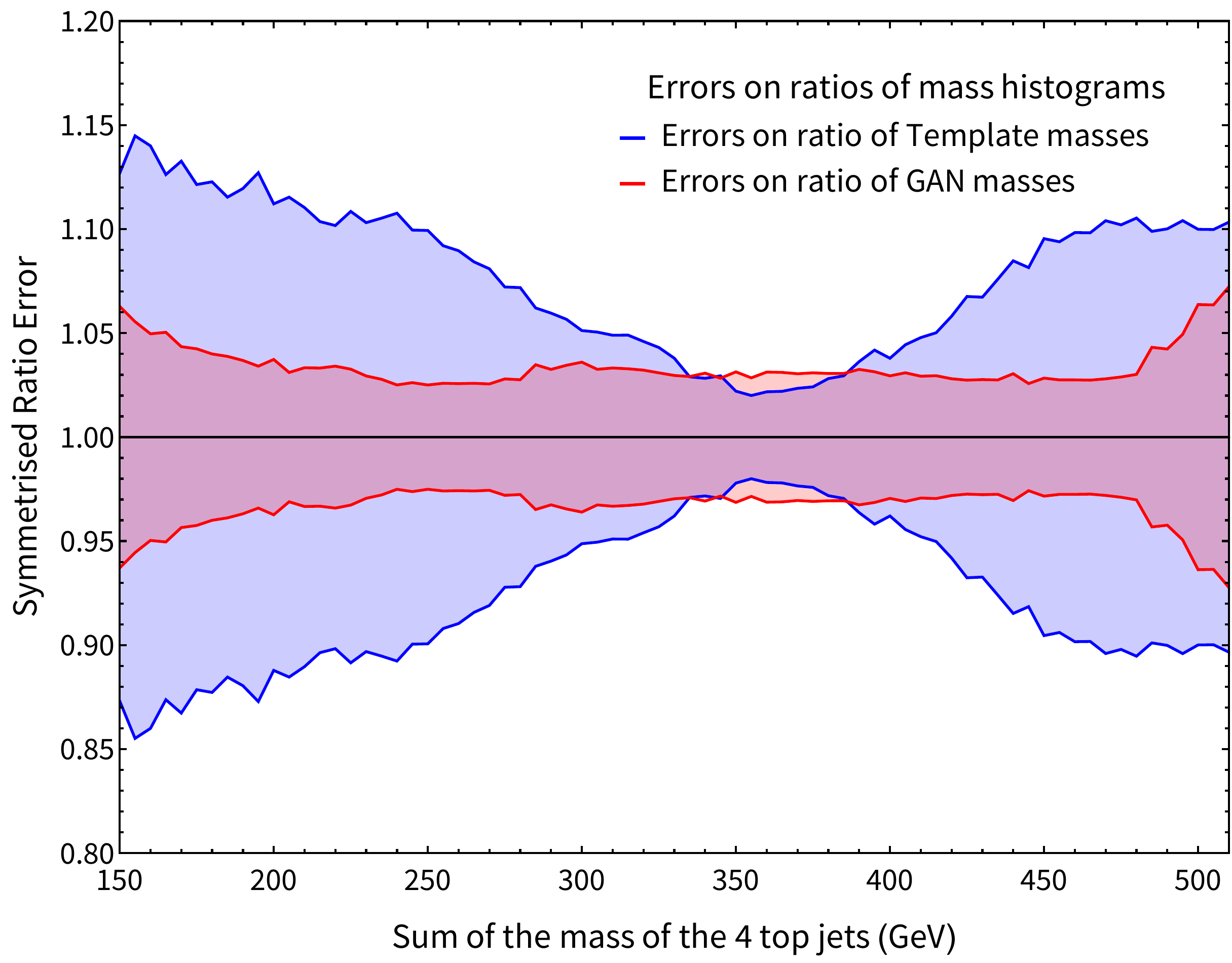}
\includegraphics[scale=0.31]{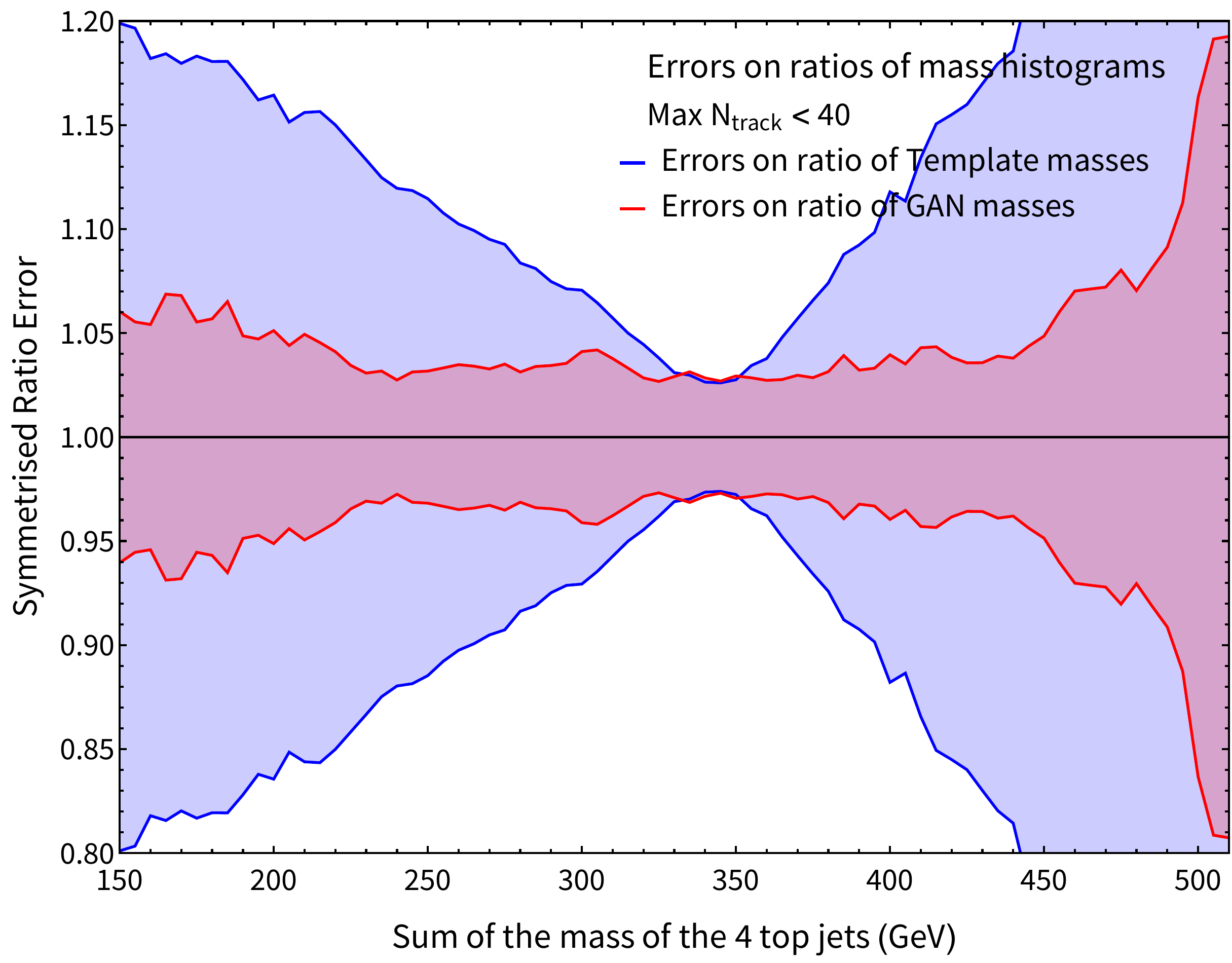}
\end{center}
\begin{center}
\includegraphics[scale=0.31]{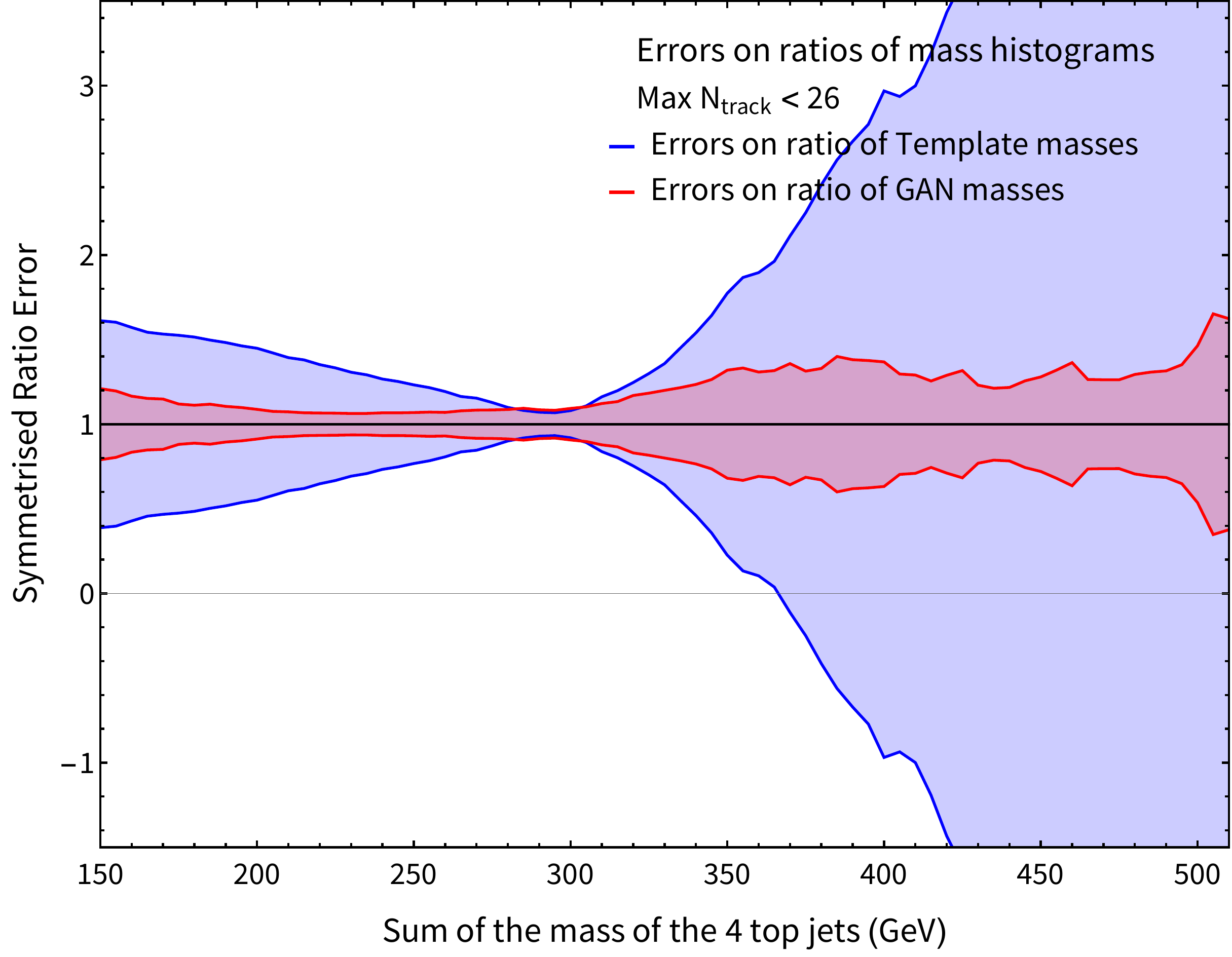}
\end{center}
\caption{Computing the ratio error of generated mass distributions ($\sum m_J$ of the masses of the top $4$ jets) to the real distribution, symmetrized by reflecting across the ideal ratio $ = 1$, with statistical uncertainties included. Inclusive error measurements are shown on the top left, for the other plots a cut is placed on $N_\text{track}$ : where for $N_\text{track} < X$ all four jets of the events considered are required to have $N_\text{track} < X$. As we reduce the efficiency of this cut, we see that the GAN's relative performance to the template distributions becomes better - expected because the GAN learns how mass changes in $N_\text{track}$.  For the top right plot, the efficiency of the cut is 50\% overall while for the bottom plot, the efficiency is about 50\% per jet (about 6\% overall).}
\label{fig:sym_ratio}
\end{figure}

%In \cref{fig:err}, we see that the GANs approximate the total distribution better than the templating method, (as we would expect, as the template data was not explicitly trained on the neutral multiplicity). 
%developed show good stability in their training as the separation power errors eventually stabilize to fixed values. Also, introducing skew into the random numbers generated for the generator part of the GAN improves the training rate of the GANs, though they all converge to the same performance. . 

The mass distributions in the validation region are presented in Fig.~\ref{fig:ratio}, in bins of jet $p_T$, $\eta$, and $N$.  The average jet mass scales approximately as $\alpha_s\times p_T\times R$ and the width of the distribution also grows with $p_T$.  Given the jet $p_T$, the jet mass should be approximately independent of $\eta$, aside from detector effects.  Gluon jets have a large jet mass than quark jets and also a higher constituent multiplicity so there is a positive correlation between $N$ and mass.

Overall, the level of agreement between the GAN and the real mass distributions is better than for the template method and the real distributions.  This is particularly true for $N$, where the real mass is shifted to lower values for low $N$ (more quark-like) and to higher values for high $N$ (more gluon-like).  The GAN is typically well within 50\% of the real distribution, while the template method can be much more than a factor of 2 off of the real distribution for low and high $N$.  This is particularly important if the quark/gluon composition is different between the control, validation, and signal regions, either by chance or because some quark-tagging is applied to suppress the QCD background in such high multiplicity final states~\cite{Sakaki:2018opq}.

The modeling of the jet mass distribution in the validation region is used to determine systematic uncertainties on the templates.  Figure~\ref{fig:sym_ratio} explicitly constructs the systematic uncertainty as the sum in quadrature of the non-closure from the validation region and the control region statistical uncertainty.  These uncertainties are computed for the $\sum_{j \in J} m_j$ distribution (sum of the masses of the top four jets in the event), which is the main observable used in the RPV SUSY search~\cite{Aad:2015lea,Aaboud:2018lpl}.  In blue we have the deviation from the exact ratio $1$ for the template mass distributions and the GAN mass distributions. The GAN outperforms the template method in both the low mass and high mass limits; also note that when placing a cut on $N_\text{track}$, the performance of the GAN becomes even more pronounced. This is particularly encouraging because we expect such RPV-SUSY signals to be quark jet dominated, with a lower multiplicity on average than a gluon jet dominated background.  For a 50\% quark jet efficiency requirement on all four jets ($N_\text{track} < 26$), the uncertainty for the GAN approach is about 20\% in the high $\sum_{i\in J} m_j$ tail while it is well over 100\% for the template appraoch.

An important part of any background estimation technique is the associated systematic uncertainty.  One of the main sources of uncertainty here is the limited size of the training set in the control region.  The authors of Ref.~\cite{Cohen:2014epa} suggested a bootstrapping technique to estimate the uncertainty by rerunning the template procedure on bootstrapped datasets.  In principle, one could do the same procedure for the GAN training, with one GAN per bootstrap dataset.  More sophisticated methods include modeling GAN weights and biases as nuisance parameters to be profiled by the data with prior distributions. 

A challenge for assessing previous GAN applications in HEP is that they have been designed to model high-dimensional feature spaces that are difficult to visualize and study~\cite{Paganini:2017hrr,Paganini:2017dwg,deOliveira:2017rwa,Chekalina:2018hxi,Carminati:2018khv,Vallecorsa:2018zco,Erdmann:2018jxd,Musella:2018rdi,Erdmann:2018kuh,ATL-SOFT-PUB-2018-001,deOliveira:2017pjk}.  The mass distribution example presented here provides a concrete testing ground to study GAN approaches where quantitative agreement can be studied and achieved using existing techniques.  While this study used only leading-order simulations of jet production, the methods are applicable more generally and can be applied to collision data from the LHC experiments.

%\textbf{Are you actually using the neutral multiplicity??}

%%%%%%%%%%%%%%%%%%%%%%%%%%
\section{Conclusions}
\label{sec:conclusions}
%%%%%%%%%%%%%%%%%%%%%%%%%%

Generative Adversarial Networks have been proposed as an alternative to histogram-based mass templates for the background estimation in LHC searches for RPV SUSY. These methods rely on the approximate QCD factorization whereby a jets type and kinematic properties are sufficient for determining the distribution of the jet mass. The neural network approaches are naturally unbinned and can be readily conditioned on multiple jet properties. In addition to using vanilla GANs for this purpose, a modification to the traditional WGAN approach has been investigated where weight clipping is replaced with drawing weights from a naturally compact set (in this case, the circle). Both the vanilla and modified WGAN approaches were able to outperform the histogram method, especially when modeling the dependence on features not used in the histogram construction. When training such generative models for physical applications, the usual limitations of the method apply such as the potential for overfitting, sensitivity to hyperparameters, and vanishing gradients slowing training down - though methods of circumnavigating these limitations have been studied in the last few years, such as using 'softer metrics' such as the Wasserstein metric. These results can be useful for enhancing the sensitivity of LHC searches to high-multiplicity final states involving many quarks and gluons and serve as a useful benchmark where GANs may have immediate benefit to the HEP community.

%In this paper, we have shown that Generative Adversarial Networks are a viable alternative to creating mass templates for searches at the LHC such as the search for RPV-SUSY. We also introduce a modification to the traditional WGAN neural network, which offers a more natural alternative to the weight clipping technique that enforces a global constant $K$ for which all the functions from the discriminator are $K$-lipschitz. Using the separation power metric, we found that the GANs were able to supercede the performance of the mass templates when including the effect of the neutral track multiplicity on the mass of the jets. 

%%%%%%%%%%%%%%%%%%%%%%%%
\acknowledgments

We would like to thank Tim Cohen, Luke de Oliveira, Mustafa Mustafa, Michela Paganini, Max Swiatloski, and Jesse Thaler for their helpful feedback on the manuscript.  This work was supported by the U.S.~Department of Energy, Office of Science under contract DE-AC02-05CH11231. 

\bibliographystyle{jhep}
\bibliography{myrefs}
\end{document}